\documentclass[pra,superscriptaddress,amsmath,amssymb,twocolumn]{revtex4}

\input{Qcircuit}
\newcommand{\braket}[2]{\langle #1 | #2 \rangle}

\newcommand{\eg}{\textit{e.g.}~}
\newcommand{\cI}{\mathcal I}
\newcommand{\Tr}{\ensuremath{\mathrm{Tr}}}
\newcommand{\cL}{\mathcal L}
\newcommand{\cH}{\mathcal H}

\usepackage{tikz}
\usetikzlibrary{decorations.markings}
\pgfdeclarelayer{edgelayer}
\pgfdeclarelayer{nodelayer}
\pgfsetlayers{edgelayer,nodelayer,main}
\tikzset{bend angle=45}
\tikzset{baseline=(current bounding box.center)}
\tikzset{column sep=2ex}
\tikzset{row sep=2ex}
\tikzstyle{none}=[inner sep=1pt]
\tikzstyle{None}=[circle, fill=white, inner sep=0pt]
\tikzstyle{box}=[draw=black, fill=white, inner sep=.3ex, rounded corners=.1ex]
\tikzstyle{dot}=[circle, draw=black, fill=black!50, inner sep=1.25pt]
\tikzstyle{cross}=[preaction={draw=white, -, line width=3pt}]
\tikzstyle{arrow}=[postaction=decorate]
\tikzstyle{thick}=[line width = 0.3mm]
\newcommand{\markat}{0.5}
\newcommand{\markwithsym}{>}
\newcommand{\markwith}{{\arrow[black]{\markwithsym}}}
\pgfkeys{/tikz/.cd, markat/.store in=\markat, markwith/.store in=\markwithsym} 
\tikzset{decoration={markings, mark=at position \markat with \markwith}}

\newcommand{\udots}{\mathinner{\mskip1mu\raise1pt\vbox{\kern7pt\hbox{.}}
   \mskip2mu\raise4pt\hbox{.}\mskip2mu\raise7pt\hbox{.}\mskip1mu}}

\begin{document}

\title{Entangled and sequential quantum protocols with dephasing}

\author{Sergio Boixo} \affiliation{Harvard University and University of Southern California} 
\author{Chris Heunen} \affiliation{Oxford University}

  \begin{abstract}
    Sequences of commuting quantum operators can be parallelized using
    entanglement. This transformation is behind some optimal quantum
    metrology protocols and recent results on quantum circuit
    complexity. We show that dephasing quantum maps in arbitrary
    dimension can also be parallelized. This implies that for general
    dephasing noise the protocol with entanglement is not more fragile
    than the corresponding sequential protocol and, conversely, the
    sequential protocol is not less effective than the entangled
    one. We derive this result using tensor networks. Furthermore, we
    only use transformations strictly valid within string diagrams in
    dagger compact closed categories. Therefore, they apply verbatim
    to other theories, such as geometric quantization and topological
    quantum field theory. This clarifies and characterizes to some
    extent the role of entanglement in general quantum theories.
 \end{abstract}

\maketitle

%\section{Introduction}
One of the main goals of quantum computation and quantum information
is the understanding of entanglement and its use to surpass the
classical bounds for some given task. Quantum metrology is a case in
point. It tries to exploit entanglement to measure
physical parameters of a system to high
precision~\cite{Helstrom1976,caves_quantum-mechanical_1981,Holevo1982,wineland_spin_1992,
braunstein_statistical_1994,bollinger_optimal_1996,huelga_improvement_1997,buzek_optimal_1999,meyer_experimental_2001,ulam-orgikh_spin_2001,fujiwara_estimation_2001,lee_quantum_2002,rudolph_quantum_2003,mitchell_super-resolving_2004,walther_broglie_2004,andre_stability_2004,luis_nonlinear_2004,bagan_quantum_2004,dunningham_sub-shot-noise-limited_2004,chiribella_efficient_2004,beltran_breaking_2005,de_burgh_quantum_2005,boixo_decoherence_2006,monras_optimal_2006,hayashi_parallel_2006,giovannetti_quantum_2006,fujiwara_strong_2006,hayashi_quantum_2006,higgins_entanglement-free_2007,shaji_qubit_2007,boixo_generalized_2007,rey_quantum-limited_2007,van_dam_optimal_2007,monras_optimal_2007,nagata_beating_2007,boixo_quantum_2008,choi_bose-einstein_2008,boixo_quantum-limited_2008,fujiwara_fibre_2008,pezze_mach-zehnder_2008,jones_magnetic_2009,appel_mesoscopic_2009,chase_magnetometry_2009,maldonado-mundo_metrological_2009,dorner_optimal_2009,lee_optimization_2009,aspachs_phase_2009,maccone_robust_2009,caves_quantum-circuit_2010,genoni_optical_2010,kacprowicz_experimental_2010,modi_role_2010,tilma_entanglement_2010,rivas_precision_2010,zwierz_general_2010,kolstrokodynacuteski_phase_2010,giovannetti_advances_2011,escher_general_2011,napolitano_interaction-based_2011,mullan_improving_2011}. 
Consider, for example, the phase estimation problem in which one is
given a phase gate $e^{-i \phi \sigma_z /2}$ as a black box and
one is to find the corresponding phase $\phi$. 
% The operator $\sigma_z$ is the Pauli $z$ matrix. 
A standard method to approach this problem is
the so-called Ramsey interferometry, where the unitary $e^{-i \phi
  \sigma_z /2}$ is applied to each subsystem of an initial quantum
state $ \left(\ket 0 + \ket 1\right)^{\otimes n} /\sqrt{2^n}$.  We are
interested in the scaling with $n$ of the uncertainty $\delta \hat
\phi$ of the estimator $\hat \phi$. In Ramsey interferometry $\delta \hat \phi \propto 1/\sqrt n$,
a scaling called the \emph{shot noise} or \emph{standard quantum limit}. In fact, as long as
the initial state remains separable and one operator is applied to each
subsystem, this bound cannot be surpassed (see supplementary material).

One way to improve this scaling is to introduce entanglement on the probe. The initial state, prior to the evolution,
is now transformed into $ \left(\ket{0\cdots 0} +\ket {1 \cdots
    1}\right)/ \sqrt 2$. This state can be achieved with an
entangling operator (as described in detail below). Applying the phase gate to
each subsystem and then a disentangling operator we obtain $  \left(\ket 0
  + e^{-i n \phi} \ket 1\right) \otimes \ket{0 \cdots 0}/\sqrt 2$. The
uncertainty scaling is now $\delta \hat \phi \propto 1/n$, which is called the Heisenberg limit~\cite{giovannetti_quantum_2006,boixo_generalized_2007}. Intuitively, entanglement in the probe
improves the measurement
sensitivity~\cite{caves_quantum-mechanical_1981}. This principle has
been corroborated experimentally on several
occasions~\cite{walther_broglie_2004,mitchell_super-resolving_2004,nagata_beating_2007}.

The scaling $\delta \hat \phi \propto 1/n$ can also be obtained in a
different way.  The final state of the first qubit in the entangled
protocol is $ \left( \ket 0 + e^{-i n \phi} \ket 1\right)/ \sqrt
2$. The same state results when acting with $n$ sequential phase gates
$\left(e^{-i \phi \sigma_z/2}\right)^n = e^{-i n \phi \sigma_z /2}$ on
the first qubit initialized to $(\ket 0 + \ket 1 )/ \sqrt 2$. Notice that,
if with a single application of the phase gate per measurement the
uncertainty is constant $\delta \hat \phi \propto 1$, then with $n$
sequential unitaries the uncertainty is $\delta(n \hat \phi) \propto
1$, giving $\delta \hat \phi \propto 1/n$. This sequential version
has been explored in frame
synchronization~\cite{rudolph_quantum_2003} and clock
synchronization~\cite{boixo_decoherence_2006,higgins_entanglement-free_2007}
between two parties.  

Given a unitary operator $U$ as a black box and an input state $\ket
\psi = \sum_j c_j \ket{e_j}$, where $\ket{e_j}$ are eigenvectors
of $U = \sum e^{i \varphi_j} \ket{e_j} \bra{e_j}$, Kitaev's phase estimation algorithm~\cite{kitaev_quantum_1995}
outputs an approximation of the phase $\varphi_j$ corresponding to
$\ket{e_j}$ with probability $|c_j|^2$. This algorithm is known to
be optimal~\cite{van_dam_optimal_2007}, and it can also be
transformed into the sequential protocol~\cite{knill_optimal_2007}. The sequentialization is
done by changing the quantum Fourier transform used in Kitaev's phase
estimation into a semi-classical Fourier
transform~\cite{griffiths_semiclassical_1996}.

The transformation between sequential and entangled protocols has also
been used to study the class QNC of quantum circuits with
polylogarithmic depth~\cite{moorenilsson:parallel}. This class
includes, for instance, standard quantum error-correction encoding and
decoding. Interestingly, if we had entangling gates with arbitrary
fan-out at our disposal, then certain important functions that require
classical circuits of logarithmic depth could be
computed by quantum circuits of constant depth~\cite{hoyer_quantum_2003}.

%\section{Perfect sequentializability}
In this paper we study the relation between entangled and sequential protocols using
\emph{tensor
  networks}~\cite{schuch_computational_2007,perez-garcia_matrix_2007,vidal_entanglement_2007,verstraete_matrix_2008,dawson_unifying_2008,cirac_renormalization_2009,evenbly_tensor_2011},
which encompass similar notations for Liouville
space~\cite{mukamel_principles_1999}, quantum
games~\cite{gutoski_toward_2007} and the so-called quantum
combs~\cite{chiribella_quantum_2008,chiribella_theoretical_2009}. We
first introduce this notation for protocols with unitary
operators. The diagrams derived are indeed very general, and they hold
verbatim for quantum maps. Therefore, we extend the relation between
entangled and sequential protocols to the presence of noise.

One reason for the generality of the diagrams used is that we only
employ transformations strictly valid within string diagrams in dagger
compact closed categories. Consequently, they apply to other quantum
theories where the maps are not tensors (for instance, they can be of
topological nature). While we refer to the literature for a formal
introduction to this subject~\cite{coeckepaquette:practisingphysicist,
  selinger:graphicallanguages, abramskycoecke:categoricalsemantics,
  joyalstreet:geometry, joyalstreet:braided}, for the purpose of this
paper it suffices to say that dagger compact closed categories capture
exactly the necessary mathematically structure that makes the
manipulations with string diagrams possible. In particular,
composition is denoted by connecting maps, there is an abstract tensor
structure between maps indicated by drawing maps in parallel, an
abstract dagger action is represented by switching the input and the output of a
map, and lines (the input/output of operations) can cross. All these
operations are \emph{compatible} between them, so the intuition gained
with tensor diagrams still applies. Reciprocally, for tensor networks
of quantum maps we will also use some adornments introduced in the
context of completely positive categories which keep track (and partly
explain) the emergence of transpose operations when carrying out certain
manipulations (see Eq.~\ref{eq:quantum_maps_adornments}).

For a given Hilbert space $\cH$ and a choice of basis $\{j\}$ the
isometry $\delta = \sum_j \ket {jj} \bra {j}$ is an entangling
operator. We denote it by
$\vcenter{\hbox{\begin{tikzpicture}[scale=0.33]
	\begin{pgfonlayer}{nodelayer}
		\node [style=none] (0) at (1, .5) {};
		\node [style=none] (1) at (-1, 0) {};
		\node [style=dot] (2) at (0, 0) {};
		\node [style=none] (3) at (1, -.5) {};
	\end{pgfonlayer}
	\begin{pgfonlayer}{edgelayer}
		\draw (1) to (2);
		\draw[out=-57,in=183] (2) to (3);
		\draw[out=57,in=177] (2) to (0);
	\end{pgfonlayer}
      \end{tikzpicture}}}$. This is a
  generalization of the following quantum circuit:
  \begin{align}
    \label{eq:1}
    \vcenter{\vspace*{1.4em}\Qcircuit@C=1em@R=.7em@!R{
      & \ctrl{1} &\qw\\
      \lstick{\ket{0}} & \targ & \qw
  }}
  \end{align}
 % The defining properties
%   satisfied by this structure %(and the category theory generalization)
%   are associativity, commutativity, the so-called Frobenius law, and
%   the isometry property $\delta^\dag \after \delta =\openone$. 
%   The latter graphically becomes
%   \mbox{$\vcenter{\hbox{\begin{tikzpicture}[scale=0.33]
% 	\begin{pgfonlayer}{nodelayer}
% 		\node [style=none] (0) at (0, 0.5) {};
% 		\node [style=none] (1) at (-2, 0) {};
% 		\node [style=dot] (2) at (-1, 0) {};
% 		\node [style=dot] (3) at (1, 0) {};
% 		\node [style=none] (4) at (2, 0) {};
% 		\node [style=none] (5) at (0, -0.5) {};
% 	\end{pgfonlayer}
% 	\begin{pgfonlayer}{edgelayer}
% 		\draw[out=237, in=357] (3) to (5.center);
% 		\draw[out=-57, in=183] (2) to (5.center);
% 		\draw[out=3, in=123] (0.center) to (3);
% 		\draw[out=57, in=177] (2) to (0.center);
% 		\draw (3) to (4.center);
% 		\draw (1.center) to (2);
% 	\end{pgfonlayer}
%       \end{tikzpicture}}} = \vcenter{\hbox{\begin{tikzpicture}[scale=0.33]
% 	\begin{pgfonlayer}{nodelayer}
% 		\node [style=none] (0) at (-1, 0) {};
% 		\node [style=none] (1) at (1, 0) {};
% 	\end{pgfonlayer}
% 	\begin{pgfonlayer}{edgelayer}
% 		\draw (0) to (1);
% 	\end{pgfonlayer}
%       \end{tikzpicture}}}$}. The operator $\delta^\dagger$ corresponds
% to a disentangling gate in quantum computation. We note that in the
% category of cobordisms, which is of relevance for quantum field
% theory, the isometry property implies that the only classical
% structure is trivial.
We now characterize commuting operators in finite-dimensional Hilbert
spaces. States are represented by the diagram
\mbox{$\vcenter{\hbox{\begin{tikzpicture}[scale=0.33, rotate=-90]
      \node [style=circle, draw=black] (state) at (0,0) {};
      \draw  (state) to (0,1) {};
    \end{tikzpicture}}}$}.
For a given state $\ket a$,  we can use the entangling operator to obtain a new operator 
\begin{align}\label{eq:abs_phase_point}
  \begin{tikzpicture}[scale=0.5, rotate=-90]
        \begin{pgfonlayer}{nodelayer}
             	\node [style=none] (2) at (0, 0) {};
        	\node [style=box,minimum size = .5cm] (3) at (0, -1) {$\;a\;$};
        	\node [style=none] (4) at (0, -2) {};
        \end{pgfonlayer}
        \begin{pgfonlayer}{edgelayer}
        	\draw (2) to (3);
        	\draw (3) to (4.center);
        \end{pgfonlayer}
 \end{tikzpicture}\quad
 : = \quad  \begin{tikzpicture}[scale=0.33, rotate=-90]
     \node [style=circle,draw=black] (0) at (-1, 2) {$a$};
     \node [style=none] (1) at (1, 2) {};
     \node [style=none] (2) at (-1, 1) {};
     \node [style=none] (3) at (1, 1) {};
     \node [style=dot] (4) at (0, 0) {};
     \node [style=none] (5) at (0, -1) {};
     \draw[looseness=1.25, bend left=315] (2.center) to (4);
     \draw (4) to (5.center);
     \draw (3.center) to (1);
     \draw (2.center) to (0);
     \draw[looseness=1.25, bend right=45] (4) to (3.center);
   \end{tikzpicture} 
 \end{align} 
which explicitly acts as
\begin{align*}%\label{eq:concr_phase_point}
  \sum_j \ket x \quad\mapsto\quad \sum_j  \braket{a}{j} \braket j x
  \ket{j}  = \bra a \left( \sum_j \ket{jj} \bra j \right) \ket x\,.
\end{align*}
Commuting operators correspond exactly to different states with the
same entangling operator or choice of basis.  These operators also
commute with the entangling operator (see supplementary material for a diagrammatic proof based on the
associative property of $\delta$.) 
\begin{align}\label{eq:associative}
  \begin{tikzpicture}[scale=0.4, rotate=-90]
        \begin{pgfonlayer}{nodelayer}
        	\node [style=none] (0) at (-1, 1) {};
        	\node [style=none] (1) at (1, 1) {};
        	\node [style=dot] (2) at (0, 0) {};
        	\node [style=box,minimum size = .5cm] (3) at (0, -1.5) {$\;a\;$};
        	\node [style=none] (4) at (0, -3) {};
        \end{pgfonlayer}
        \begin{pgfonlayer}{edgelayer}
        	\draw (2) to (3);
        	\draw[bend right=45, looseness=1.5] (2) to (1.center);
        	\draw (3) to (4.center);
        	\draw[bend right=45, looseness=1.5] (0.center) to (2);
        \end{pgfonlayer}
 \end{tikzpicture}
 \quad & = \quad
 \begin{tikzpicture}[scale=0.4, rotate=-90]
        \begin{pgfonlayer}{nodelayer}
        	\node [style=none] (0) at (-1, 2) {};
        	\node [style=none] (1) at (1, 2) {};
        	\node [style=box,minimum size = .5cm] (2) at (-1, 1) {$\;a\;$};
        	\node [style=none] (3) at (1, 1) {};
        	\node [style=dot] (4) at (0, 0) {};
        	\node [style=none] (5) at (0, -1) {};
        \end{pgfonlayer}
        \begin{pgfonlayer}{edgelayer}
        	\draw (4) to (5.center);
        	\draw[bend right=45, looseness=1.50] (2.center) to (4);
        	\draw[bend right=45, looseness=1.25] (4) to (3.center);
        	\draw (1.center) to (3.center);
        	\draw (0.center) to (2);
        \end{pgfonlayer}
 \end{tikzpicture} \\
 % \sum_{j,k} \ket{kk} (\bra{k} a \ket{j} \otimes \ket{j}) \bra{j} =
  \sum_j \ket{jj} \bra{j} a \ket{j} \bra{j}
  \quad & = \quad 
  \sum_j (a \otimes 1) \ket{jj} \bra{j}
\end{align}
The physical interpretation is that there is no difference whatsoever
between applying a commuting gate first and then the
entangling operator or first entangling and then applying the gate. 

Writing \mbox{$\vcenter{\hbox{\begin{tikzpicture}[scale=0.33, rotate=-90]
   \node [style=none] (0) at (0, 1) {}; \node [style=dot] (1) at (0, 0) {}; \draw (1) to (0) {};
\end{tikzpicture}}}$} for the (unnormalized) equal superposition state
$\sum \ket{j}$ for the given choice of basis, and
\mbox{$\vcenter{\hbox{\begin{tikzpicture}[scale=0.33, rotate=90] 
   \node [style=none] (0) at (0, 1) {}; \node [style=dot] (1) at (0, 0) {}; \draw (1) to (0) {};
\end{tikzpicture}}}$} for its adjoint $\sum \bra{j}$, we obtain equivalence
between the following two diagrams for any set of commuting operators $\{f_j\}$: 
\begin{align}\label{eq:parallelprotocol}
  \vcenter{\hbox{\begin{tikzpicture}[scale=0.4]
	\begin{pgfonlayer}{nodelayer}
		\node [style=none] (0) at (-1, 2) {};
		\node [style=box] (1) at (0, 2) {$\;\;f_n\;\;$};
		\node [style=none] (2) at (1, 2) {};
		\node [style=none] (3) at (-3, 1) {};
		\node [style=dot] (4) at (-2, 1) {};
		\node [style=dot] (5) at (2, 1) {};
		\node [style=none] (6) at (3, 1) {};
		\node [style=none] (7) at (-3.5, 0.5) {$\udots$};
		\node [style=none] (8) at (3.5, 0.8) {$\ddots$};
		\node [style=none] (8a) at (0, -1) {$\vdots$};
		\node [style=none] (9) at (-4, 0) {};
		\node [style=none] (10) at (-1, 0) {};
		\node [style=box] (11) at (0, 0) {$f_{n-1}$};
		\node [style=none] (12) at (1, 0) {};
		\node [style=none] (13) at (4, 0) {};
		\node [style=dot] (14) at (-5, -1) {};
		\node [style=dot] (15) at (5, -1) {};
		\node [style=none] (16) at (-4, -2) {};
		\node [style=none] (17) at (-1, -2) {};
		\node [style=box] (18) at (0, -2) {$\;\;f_2\;\;$};
		\node [style=none] (19) at (1, -2) {};
		\node [style=none] (20) at (4, -2) {};
		\node [style=dot] (21) at (-7, -2.25) {};
		\node [style=dot] (22) at (-6, -2.25) {};
		\node [style=dot] (23) at (6, -2.25) {};
		\node [style=dot] (24) at (7, -2.25) {};
		\node [style=none] (25) at (-5, -3.5) {};
		\node [style=none] (26) at (-1, -3.5) {};
		\node [style=box] (27) at (0, -3.5) {$\;\;f_1\;\;$};
		\node [style=none] (28) at (1, -3.5) {};
		\node [style=none] (29) at (5, -3.5) {};
	\end{pgfonlayer}
	\begin{pgfonlayer}{edgelayer}
		\draw (25.center) to (26.center);
		\draw[bend left=45] (22) to (14);
		\draw (16.center) to (17.center);
		\draw[bend left=45, looseness=1.25] (13.center) to (15);
		\draw[bend right=45] (29.center) to (23);
		\draw (21) to (22);
		\draw (19.center) to (20.center);
		\draw[bend left=45, looseness=1.25] (2.center) to (5);
		\draw[bend left=45, looseness=1.25] (15) to (23);
		\draw (26.center) to (28.center);
		\draw[bend right=45, looseness=1.25] (20.center) to (15);
		\draw[bend left=315] (22) to (25.center);
		\draw[bend left=45, looseness=1.25] (14) to (9.center);
		\draw[bend right=45, looseness=1.25] (4) to (10.center);
		\draw (28.center) to (29.center);
		\draw (3.center) to (4);
		\draw (10.center) to (12.center);
		\draw (23) to (24);
		\draw (17.center) to (19.center);
		\draw (5) to (6.center);
		\draw[bend left=45, looseness=1.25] (4) to (0.center);
		\draw (0.center) to (2.center);
		\draw[bend right=45, looseness=1.25] (14) to (16.center);
		\draw[bend right=45, looseness=1.25] (12.center) to (5);
	\end{pgfonlayer}
      \end{tikzpicture}}} 
\end{align}
\begin{align}\label{eq:sequentialprotocol}
\vcenter{\hbox{\begin{tikzpicture}[scale=0.75]
	\begin{pgfonlayer}{nodelayer}
		\node [style=dot] (0) at (-4.5, 0) {};
		\node [style=box] (1) at (-3, 0) {$\;f_{\pi(1)}\;$};
		\node [style=box] (2) at (-1.5, 0) {$\;f_{\pi(2)}\;$};
		\node [style=none] (3) at (-.5, 0) {};
		\node [style=none] (4) at (-0.05, -0.05) {$\cdots$};
		\node [style=none] (5) at (.4, 0) {};
		\node [style=box] (6) at (1.5, 0) {$f_{\pi(n-1)}$};
		\node [style=box] (7) at (3.1, 0) {$\,f_{\pi(n)}\,$};
		\node [style=dot] (8) at (4.5, 0) {};
	\end{pgfonlayer}
	\begin{pgfonlayer}{edgelayer}
		\draw (2) to (3.center);
		\draw (5.center) to (6);
		\draw (7) to (8);
		\draw (0) to (1);
		\draw (6) to (7);
		\draw (1) to (2);
	\end{pgfonlayer}
      \end{tikzpicture}}}
\end{align}
for any permutation $\pi \in
S(n)$. Diagram~\eqref{eq:parallelprotocol} corresponds to the 
entangled protocol in three steps: entangling gate, commuting gate
acting on each component, disentangling
gate. Diagram~\eqref{eq:sequentialprotocol} clearly corresponds to the
sequential protocol.  The equivalence follows diagrammatically because
every commuting operator $f_j$ can be moved to the beginning of the
diagram, and the remaining loops can be cancelled by the isometry
property \mbox{$\vcenter{\hbox{\begin{tikzpicture}[scale=0.33]
	\begin{pgfonlayer}{nodelayer}
		\node [style=none] (0) at (0, 0.5) {};
		\node [style=none] (1) at (-2, 0) {};
		\node [style=dot] (2) at (-1, 0) {};
		\node [style=dot] (3) at (1, 0) {};
		\node [style=none] (4) at (2, 0) {};
		\node [style=none] (5) at (0, -0.5) {};
	\end{pgfonlayer}
	\begin{pgfonlayer}{edgelayer}
		\draw[out=237, in=357] (3) to (5.center);
		\draw[out=-57, in=183] (2) to (5.center);
		\draw[out=3, in=123] (0.center) to (3);
		\draw[out=57, in=177] (2) to (0.center);
		\draw (3) to (4.center);
		\draw (1.center) to (2);
	\end{pgfonlayer}
      \end{tikzpicture}}} = \vcenter{\hbox{\begin{tikzpicture}[scale=0.33]
	\begin{pgfonlayer}{nodelayer}
		\node [style=none] (0) at (-1, 0) {};
		\node [style=none] (1) at (1, 0) {};
	\end{pgfonlayer}
	\begin{pgfonlayer}{edgelayer}
		\draw (0) to (1);
	\end{pgfonlayer}
      \end{tikzpicture}}}$} (See supplementary material).
Explicitly, the main step is as follows:
\begin{equation}\label{eq:useisometry}
  \sum_{k,l} \ket{k} \bra{kk} 1 \otimes f_j \ket{ll} \bra{l}
%   = \sum_k \bra{k} \ket{k} f_j \bra{k} \ket{k}
  = f_j
%  = \sum_k \ket{k}\bra{k} f_j
  = \big (\sum_{k,l} \ket{k} \braket{kk}{ll} \bra{l} \big ) \circ f_j,
\end{equation}
where $f_j$ can be represented as in Eq.~\ref{eq:abs_phase_point}.
Note that this equivalence also establishes the commutativity of the
operators among themselves. 

We have shown that the operators defined by the diagram of
Eq.~\eqref{eq:abs_phase_point} commute with the entangling gate. Using
the (unnormalized) equal superposition state, we can prove the
converse. Given an operator $a$ that commutes with an entangling
operator, it is easy to see that the following diagram defines the
corresponding state (see supplementary material):
\begin{align}\label{eq:abs_operator_state}
  \begin{tikzpicture}[scale=0.33, rotate=-90]
    \begin{pgfonlayer}{nodelayer}
        \node [style=circle,draw=black] (3) at (0, 0) {$a$};
      \node [style=none] (4) at (0, 2) {};
    \end{pgfonlayer}
    \begin{pgfonlayer}{edgelayer}
     \draw (3) to (4.center);
    \end{pgfonlayer}
  \end{tikzpicture}
 \quad := \quad
  \begin{tikzpicture}[scale=0.33, rotate=-90]
    \begin{pgfonlayer}{nodelayer}
      \node [style=none] (2) at (0, 0) {};
      \node [style=box,minimum size = .5cm] (3) at (0, -1.5) {$\;a\;$};
      \node [style=dot] (4) at (0, -3) {};
    \end{pgfonlayer}
    \begin{pgfonlayer}{edgelayer}
      \draw (2) to (3); 
      \draw (3) to (4.center);
    \end{pgfonlayer}
  \end{tikzpicture}
\end{align}
Explicitly, the state $\ket{a} = \sum_j a \ket{j}$ and the operator
$a$ are related by Eq.~\eqref{eq:abs_phase_point}.

We have given a complete diagrammatic characterization of all the
commuting operators and reviewed the known
equivalence of the corresponding entangled and sequential protocols.
We now extend this equivalence to quantum maps with general dephasing
(a similar question has been addressed before for maps on
qubits~\cite{boixo_decoherence_2006}). 
To distinguish $\cH$ from its dual space $\cH^*$, we annotate the
corresponding wires in diagrams with opposite arrows. Thus 
operators $c$ from $\cH_A$ to $\cH_B$ come in four variations:
the original, the transpose $c^*$, the adjoint $c^\dagger$, and the conjugate $c_* =
(c^\dagger)^*$. 
\begin{align}
     & \begin{tikzpicture}[scale=1,rotate=-90]
      \begin{pgfonlayer}{nodelayer}
     \node  (1) at (0,-1) {$\cH_A$};
     \node  (2)  [style=circle,draw=black] at (0,0) {$c\phantom{^\dag}$};
     \node  (3) at (0,1) {$\cH_B$};
     \end{pgfonlayer}
     \begin{pgfonlayer}{edgelayer}
     \draw [arrow] (1) to (2) {};
     \draw [arrow] (2) to (3) {};
     \end{pgfonlayer}
   \end{tikzpicture}  &
   & \begin{tikzpicture}[scale=1,rotate=-90]
      \begin{pgfonlayer}{nodelayer}
     \node  (1) at (0,-1) {$\cH_B^*$};
     \node  (2)  [style=circle,draw=black] at (0,0) {$c^*$};
     \node  (3) at (0,1) {$\cH_A^*$};
     \end{pgfonlayer}
     \begin{pgfonlayer}{edgelayer}
     \draw [arrow] (3) to (2) {};
     \draw [arrow] (2) to (1) {};
     \end{pgfonlayer}
   \end{tikzpicture}  \nonumber \\
   & \begin{tikzpicture}[scale=1,rotate=-90]
      \begin{pgfonlayer}{nodelayer}
     \node  (1) at (0,-1) {$\cH_B$};
     \node  (2)  [style=circle,draw=black] at (0,0) {$c^\dagger$};
     \node  (3) at (0,1) {$\cH_A$};
     \end{pgfonlayer}
     \begin{pgfonlayer}{edgelayer}
     \draw [arrow] (1) to (2) {};
     \draw [arrow] (2) to (3) {};
     \end{pgfonlayer}
   \end{tikzpicture} &
   & \begin{tikzpicture}[scale=1,rotate=-90]
      \begin{pgfonlayer}{nodelayer}
     \node  (1) at (0,-1) {$\cH_A^*$};
     \node  (2)  [style=circle,draw=black] at (0,0) {$c_*$};
     \node  (3) at (0,1) {$\cH_B^*$};
     \end{pgfonlayer}[
     \begin{pgfonlayer}{edgelayer}
     \draw [arrow] (3) to (2) {};
     \draw [arrow] (2) to (1) {};
     \end{pgfonlayer}
   \end{tikzpicture} \label{eq:quantum_maps_adornments}
\end{align}
We can now use the notation
$\vcenter{\smash{\hbox{\begin{tikzpicture}[scale=0.15, rotate=-90]
    \node  (1) at (1,0) {};
     \node  (10)  at (4, 0) {};
    \draw [arrow,looseness=1, markwith={>}, bend left=90] (10) to (1) {};
  \end{tikzpicture}}}}$ for the (entangled) state $ \sum_j \ket{\tilde{j} j}\in \cH^* \otimes \cH$, and
$\vcenter{\smash{\hbox{\begin{tikzpicture}[scale=0.15, rotate=90] 
    \node  (1) at (1,0) {};
     \node  (10)  at (4, 0) {};
    \draw [arrow,looseness=1, markwith={<}, bend left=90] (10) to (1) {};
  \end{tikzpicture}}}}$ for its adjoint.
% In the diagrammatic notation (and other contexts), it is convenient to
% relate a density operator $\rho$ with its matrix $\rho \equiv \openone
% \otimes \rho \sum \ket{\tilde j j} = \rho^* \otimes \openone \sum
% \ket{\tilde j j}$. The fact that the density operator gets transposed
% when moving from one side to the other of the maximally entangled state
% is captured by the \emph{arrow} adornment that keeps track of the
% difference between a Hilbert space and its dual. At a fundamental
% level this necessity is a symptom of the fact that a Hilbert space is
% isometrically isomorphic to its dual but, categorically speaking, this
% correspondence is not part of a \emph{natural transformation}.

Now, let $\mathcal A(\rho) = \sum_t a_t \cdot \rho \cdot a_t^\dagger$
be a completely positive map with Kraus operators $\{a_t\}$. Define
the operator $a = \sum_t \ket t \otimes a_t$ (where the states $\ket
t$ live in a environment Hilbert space $\cH_E$). Its diagrammatic
representation is
\begin{align}
     \begin{tikzpicture}[scale=1, rotate=-90]
      \begin{pgfonlayer}{nodelayer}
     \node  (1) at (0,-1) {};
     \node  (2)  [style=circle,draw=black,thick] at (0,0) {$\mathcal A$};
     \node  (3) at (0,1) {};
     \end{pgfonlayer}
     \begin{pgfonlayer}{edgelayer}
     \draw [thick] (1) to (2) {};
     \draw [thick] (2) to (3) {};
     \end{pgfonlayer}
    \end{tikzpicture}
 =
  \begin{tikzpicture}[rotate=-90]
	\begin{pgfonlayer}{nodelayer}
          \node [style=none] (0) at (-1, 1) {$\cH_B^*$};
          \node [style=none] (1) at (1, 1) {$\cH_B$};
          \node [style=none] (2) at (-1, 0) {};
          \node [style=none] (2b) at (-1, -0.3) {};
          \node [style=none] (3) at (-0.5, 0) {};
          \node [style=none] (4) at (0.5, 0) {};
          \node [style=none] (5) at (1, 0) {};
          \node [style=none] (5b) at (1, 0.2) {};
          \node [style=none] (6) at (-1, -1) {$\cH_A^*$};
          \node [style=none] (7) at (1, -1) {$\cH_A$};
          \node [style=none] at (0, 0.9) {$\cH_E$};
          \node [style=circle,draw=black,fill=white] at (-.75,0) {$a_*$};
          \node [style=circle,draw=black,fill=white] at (.75,0) {$\;a\;$};
	\end{pgfonlayer}
	\begin{pgfonlayer}{edgelayer}
          \draw[arrow] (0) to (2);
          \draw[arrow, looseness=1.75, bend right=90] (4) to (3);
          \draw[arrow] (2b) to (6);
          \draw[arrow] (7) to (5);
          \draw[arrow] (5b) to (1);
	\end{pgfonlayer}
      \end{tikzpicture}
\end{align}
The second diagram corresponds to the Stinespring representation of $\mathcal
A$. All completely positive operators can be put in this form. The
thick first diagram is defined as a more concise representation. 
Positive states $B = b^\dagger b$ also have this form, with trivial input space
$\cH_A = \mathbb{C}$. 
% \[
%   \begin{tikzpicture}[scale=0.4]
%     \begin{pgfonlayer}{nodelayer}
%     \node (0) at (-3,2) {};
%     \node [style=circle,draw=black] (1) at (-1.5,0) {$b^*$};
%     \node [style=circle,draw=black,minimum size = .75cm] (2) at (1.5,0) {$b^\dagger$};
%     \node (3) at (3,2) {};
%   \end{pgfonlayer}
%        \begin{pgfonlayer}{edgelayer}
%          \draw [arrow,looseness=1.25, bend right=25] (0) to (1) {};
%          \draw [arrow,looseness=1.25, bend right=15] (1) to (2) {};
%          \draw [arrow,looseness=1.25, bend right=25] (2) to (3) {};
%         \end{pgfonlayer}
%   \end{tikzpicture}
% \]

Every operator between Hilbert spaces can be lifted to a completely positive quantum map. The quantum map obtained by lifting the entangling operator $\delta$ for a given choice of basis has the form 
\begin{align}\label{eq:cp_classical_structure}
    \begin{tikzpicture}[scale=0.4, rotate=-90]
     \node [style=none] (0) at (-1, 2) {};
     \node [style=none] (1) at (1, 2) {};
     \node [style=none] (2) at (-1, 1) {};
     \node [style=none] (3) at (1, 1) {};
     \node [style=dot] (4) at (0, 0) {};
     \node [style=none] (5) at (0, -1) {};
     \draw[looseness=1.25, bend left=315]  (2.center) to (4);
     \draw [arrow] (4) to (5.center);
     \draw [arrow] (1) to (3.center);
     \draw [arrow] (0) to (2.center);
     \draw[looseness=1.25, bend right=315] (3.center) to (4);
    \node [style=none] (10)  at (2, 2) {};
     \node [style=none] (11) at (4, 2) {};
     \node [style=none] (12) at (2, 1) {};
     \node [style=none] (13) at (4, 1) {};
     \node [style=dot] (14) at (3, 0) {};
     \node [style=none] (15) at (3, -1) {};
     \draw[looseness=1.25, bend left=315] (12.center) to (14);
     \draw [arrow] (15.center)to (14);
     \draw [arrow] (13.center) to (11);
     \draw [arrow] (12.center) to (10);
     \draw[looseness=1.25, bend right=45] (14) to (13.center);
   \end{tikzpicture} 
   \quad = \quad    
   \begin{tikzpicture}[scale=0.4, rotate=-90]
     \node [style=none] (0) at (-1, 2) {};
     \node [style=none] (1) at (1, 2) {};
     \node [style=none] (2) at (-1, 1) {};
     \node [style=none] (3) at (1, 1) {};
     \node [style=dot, thick] (4) at (0, 0) {};
     \node [style=none] (5) at (0, -1) {};
     \draw[thick,looseness=1.25, bend left=315]  (2.center) to (4);
     \draw [thick] (4) to (5.center);
     \draw [thick] (1) to (3.center);
     \draw [thick] (0) to (2.center);
     \draw[thick,looseness=1.25, bend right=315] (3.center) to (4);
   \end{tikzpicture}
 \end{align}
For any positive state and a choice of basis we can then characterize
all commuting quantum maps in that basis. They are obtained exactly as
in the diagrams of Eq.~\eqref{eq:abs_phase_point}, and take the
form
\begin{align}\label{eq:phase_point}
    \begin{tikzpicture}[scale=0.4, rotate=-90]
      \begin{pgfonlayer}{nodelayer}
     \node [style=none] (0) at (-1, 4) {};
     \node [style=circle,draw=black] (1) at (1,2) {$b_*$};
     \node [style=none] (2) at (-1, 1) {};
     \node [style=none] (3) at (1, 1) {};
     \node [style=dot] (4) at (0, 0) {};
     \node [style=none] (5) at (0, -1) {};
     \node [style=circle,draw=black,minimum size = .75cm]  (10)  at (3, 2) {$\;b\;$};
     \node [style=none] (11) at (5, 4) {};
     \node [style=none] (12) at (3, 1) {};
     \node [style=none] (13) at (5, 1) {};
     \node [style=dot] (14) at (4, 0) {};
     \node [style=none] (15) at (4, -1) {};
     \end{pgfonlayer}
     \begin{pgfonlayer}{edgelayer}
     \draw[looseness=1.25, bend left=315]  (2.center) to (4);
     \draw [arrow] (4) to (5.center);
     \draw (1) to (3.center);
     \draw [arrow] (0) to (2.center);
     \draw [arrow, looseness=1.25, bend right=315] (3.center) to (4);
     \draw [arrow, markwith={<}, looseness=1.25, bend left=315] (12.center) to (14);
     \draw [arrow] (15.center)to (14);
     \draw [arrow] (13.center) to (11);
     \draw (12.center) to (10);
     \draw[looseness=1.25, bend right=45] (14) to (13.center);
     \draw [arrow,looseness=1, bend right=90] (10) to (1) {};
     \end{pgfonlayer}
   \end{tikzpicture}
   \quad = \quad 
   \begin{tikzpicture}[scale=1, rotate=-90]
      \begin{pgfonlayer}{nodelayer}
     \node  (1) at (0,-1) {};
     \node  (2)  [style=box,draw=black,thick] at (0,0) {$\;\;\mathcal B\;\;$};
     \node  (3) at (0,1) {};
     \end{pgfonlayer}
     \begin{pgfonlayer}{edgelayer}
     \draw [thick] (1) to (2) {};
     \draw [thick] (2) to (3) {};
     \end{pgfonlayer}
    \end{tikzpicture}
\end{align}
Notice that the diagrammatic proofs (as in the supplementary material)
stay exactly the same, whereas the direct computations as in
Eq.~\eqref{eq:useisometry} become rather more involved.

All the diagrams introduced for unitary operators that commute with
the entangling operator apply verbatim to quantum maps. Indeed, the
quantum maps constructed through an entangling map commute and all
maps commuting with an entangling map are of this particular
form. Further, the equivalence between the entangled and sequential
protocols is upheld.

We now give explicit equations for quantum maps that commute with
an entangling map. From the defining diagram
Eq.~\eqref{eq:phase_point} we see that the quantum map $\mathcal B$
corresponding to a positive operator $B$ acts as
\[
  \mathcal B(\rho) =  B^* \circ \rho\;,
\]
where $\circ$ is the Schur product in the basis of the entangling
map. All quantum maps of that form are completely positive (already
shown diagrammatically), but in addition we want them to be trace
preserving, which imposes
\begin{align}\label{eq:trace_condition}
  B_{jj} = 1\;.
\end{align}
The positive operator $B$ has a spectral decomposition
\[
  B = \sum_s \ket{\chi_s} \bra {\chi_s} \;. 
\]
This gives the Kraus decomposition of $\mathcal B(\rho) = \sum_s b_s \cdot \rho \cdot b_s^\dagger $, with Kraus operators
\[
  b_s = \sum_j \braket{\chi_s}{j} \;\ket j \bra j\;,
\]
Notice that in this representation the commuting quantum
maps also have commuting Kraus operators.

For unitary phase maps the matrix $B$ must have the form
\[
  (B_{\rm phases})_{jk} = e^{-i (\phi_j - \phi_k) } \ket j \bra k 
\]
for any choice of phases $\{\phi_j\}$.
Maps with \emph{dephasing} with respect
to the basis of the entangling map are more interesting. The
noise must be dephasing because
it does not alter the populations, due to Eq.~\eqref{eq:trace_condition}:
\[
  (\mathcal B(\rho))_{jj} = B_{jj} \rho_{jj} = \rho_{jj}\;.
\]
For the qubit case (i.e.~when the dimension of the Hilbert space is
two), the positive matrix $B$ has the form
\[
  \left(
    \begin{array}{cc}
      1 & e^{- \gamma - i \phi} \\
      e^{-\gamma + i \phi} & 1
    \end{array}\right)\;,
\]
where $\phi$ is a phase and $\gamma \ge 0$ parametrizes the
dephasing. This quantum channel is
\[
  e^{-i \phi \sigma_z /2} \cdot \left( (1-p)\; \rho + p \;\sigma_z \rho \sigma_z \right ) \cdot e^{-i \phi \sigma_z/2}\;,
\]
where $p = \frac {1-e^{- \gamma}} 2$ can be interpreted as the probability of a random
phase flip.

% We also note that the standard parallel phase estimation protocol with
% qubits can be sequentialized even with damping noise (and not only
% dephasing noise)~\cite{boixo_decoherence_2006}. The basic reason is
% that the standard parallel protocol measures on the $X$ - $Y$ plane of
% the Bloch sphere, and therefore does not ``see'' the
% damping. Nevertheless, in accordance with the results presented here,
% the prepared state is actually different.

We can also write explicitly a more generic map that is fully
sequentializable. We do that by defining appropriate (unnormalized) vectors
$\ket{\chi_s}$ for the positive matrix $B = \sum_s
\ket{\chi_s}\bra{\chi_s}$. These vectors should be orthogonal and for
the trace preserving condition should obey
\begin{align}
  B_{jj} = \sum_s \braket{j}{\chi_s} \braket{\chi_s}{j} = \sum_s |(\chi_s) _j|^2 = 1\;.
\end{align}
Denote the $n$th roots of unity by
\begin{align}
  \omega_j = e^{-i 2 \pi j/n}\;,
\end{align}
where $n$ is the dimension of the underlying Hilbert space. Define
\begin{align}
  \ket{\chi_s} = \sqrt{r_s} \sum_j e^{-i \phi_j} \omega^s_j \ket j
\end{align}
for arbitrary phases $\phi_j$ and (dephasing) positive constants $r_j$ such that $\sum_j r_j = 1$. Then
\begin{align}
  \braket{\chi_s}{\chi_t} = \sqrt{r_sr_t} \sum_j \omega^{t-s}_j =  \sqrt{r_sr_t}n \delta_{s,t}\;,
\end{align}
which makes the vectors orthogonal. Further,
\begin{align}
   \sum_s |(\chi_s) _j|^2 = \sum_s r_s = 1\;,
\end{align}
which makes the corresponding completely positive map trace preserving. The Kraus operators of this map are
\begin{align}
  b_s = \sqrt{r_s} \sum_j e^{-i (\phi_j + 2 \pi j s /n)} \ket j \bra j\;.
\end{align}
The map without dephasing (a pure rotation) is recovered by choosing $r_s = \delta_{s,0}$.

All diagrams we have used can equally well be interpreted in
(dagger compact closed) categories other than tensor networks of
Hilbert spaces.  In the context of category theory, the morphism
corresponding to the entangling gate $\delta$ is called a
\emph{classical structure}, because it abstracts the process of
copying classical information. It also abstracts a choice of
orthonormal basis~\cite{coeckepavlovicvicary:bases,abramskyheunen:hstar}. The 
corresponding ``equal superposition'' state
\mbox{$\vcenter{\hbox{\begin{tikzpicture}[scale=0.33, rotate=-90] \node
        [style=none] (0) at (0, 1) {}; \node [style=dot] (1) at (0, 0)
        {}; \draw (1) to (0) {};
              \end{tikzpicture}}}$} 
is the unique unit of the classical structure. Sequentialization and
parallelization of commuting morphisms are consequences of the
\emph{generalized spider theorem}~\cite{coeckeduncan:observables},
which is deduced by similar diagrammatic manipulations. Further, a
dagger compact closed category with completely positive morphisms can
be constructed from any dagger compact closed
category~\cite{selinger:completelypositive}, giving the corresponding
abstract ``dephasing morphisms'' that commute with the classical
structure. See~\cite{heunenboixo:cscp} for more information.
Therefore our results have consequences in other theories.
For example, conformal or topological quantum field
theories can be formulated in terms of
dagger compact closed categories of cobordisms~\cite{kock:frobenius,
baez:quandaries, atiyah:tqft, segal:cft}. Similarly, geometric
quantization can be formulated as concerning classical structures in the
dagger compact category of symplectic manifolds and canonical
relations~\cite{landsman:quantization}. Our results, giving classes
of sequentializable configurations, become interesting in those
settings when reading sequentializability as the ability to ``trade
entanglement for time''.  

% We have derived the equivalence between the entangled and sequential
% protocols using only the abstract properties of the classical
% structures. These properties are associativity, commutativity,
% isometry, and the Frobenius equation. For bounded linear operators
% between (pure states in) finite-dimensional Hilbert spaces, it is
% known that these properties characterize uniquely the entangling
% gates~\cite{coeckepavlovicvicary:bases,abramskyheunen:hstar}. A
% natural question is if there are quantum maps (between mixed states),
% other than the ones corresponding to entangling maps, with the same
% properties. This would lead to sequentializable or parallelizable
% protocols different from the ones already covered. It turns out not to
% be the case. Because a classical structure is an isometry, it obeys
% \begin{align}
%   \Tr \;(\delta(\ket \psi \! \bra \psi))^2  = \Tr \ket \psi \! \bra \psi =1\;.
% \end{align}
% Now, because a quantum map preserves trace, this implies that a
% trace preserving classical structure defines a linear operator
% between pure states, and, by the result referenced above, must
% correspond to an entangling gate.

%\section*{Conclusions} 
In summary, we have shown that quantum maps composed of phase gates and general
dephasing are exactly the maps that commute with the entangling
map. This implies that the sequential and quantum protocols are
equivalent. In particular, they are equally sensitive to noise and
have exactly the same responsiveness for quantum metrology. The entangled
(parallel) protocol is better suited for rapidly changing signals, but
is technically challenging. The diagrammatical derivation we presented
has advantages over direct computations: it is completely general and,
though perhaps unfamiliar, arguably simpler. Also, the abstract point
of view sheds some light on the fundamental structures of some quantum
protocols and applies to other theories.

\paragraph*{Acknowledgement.}

Much of this work has been done while both authors were at the
Institute for Quantum Information at the California Institute of
Technology. We thank Peter Selinger for pointing
out~\cite{mccurdyselinger:basicdaggercompactcategories}, and David
P{\'e}rez Garc{\'i}a, Robert K{\"o}nig, Peter Love and Spiros Michalakis for discussions. 
SB acknowledges support from FIS2008-01236 and from Defense Advanced
Research Projects Agency award N66001-09-1-2101.
CH was supported by the Netherlands Organisation for Scientific
Research (NWO).

\appendix
\section{Bound on the Fisher information scaling for separable states}
\label{sec:fisher_bound}

The quantum Cram\'er-Rao inequality~\cite{braunstein_statistical_1994}
states that
\begin{align}
  \delta \hat \phi \ge \frac 1 {\sqrt {\cI_n}},
\end{align}
where $\cI_n$ is the (quantum) Fisher information corresponding to the
evolution of the quantum state parametrized by $\phi$. This appendix
proves $\cI_n \le n \cI_{\rm bound}$ for separable states, where 
$\cI_{\rm bound}$ is a bound on the Fisher information for one
system. This is a well known fact, which we include here for
completeness (see \eg~\cite{boixo_quantum-limited_2008} for a
different proof). We also note that for subsystems evolving with 
Hamiltonian $h$ (so the total Hamiltonian is $\sum h$) there exists a
bound $\cI_{\rm bound} \le \|
h\|$~\cite{giovannetti_quantum_2006,boixo_generalized_2007}.

Consider a quantum state $\rho$, implicitly parametrized by $\phi$. The quantum Fisher information for the state $\rho$ is defined as
\[
  \cI_n = \Tr \rho \cL_n^2
\]
where the symmetric logarithmic derivative $\cL$ is the Hermitian operator implicitly defined by the equation
\[
  \frac 1 2 ( \cL_n \rho + \rho \cL_n) = \frac {\partial \rho}{\partial \phi}.
\]
If follows directly from this definition that
\begin{align}\label{eq:sld_zero_average}
  \Tr \rho \cL = \Tr  \frac 1 2 ( \cL_n \rho + \rho \cL_n) =  \Tr \frac {\partial \rho}{\partial \phi} =  \frac {\partial \; \Tr \rho}{\partial \phi} = 0.
\end{align}

Now assume that $\rho$ is a product quantum state
$\rho_p = \rho^{(1)} \otimes \cdots \otimes \rho^{(n)}$.
It is easy to check directly that if $\cL^{(j)}$ is the symmetric
logarithmic derivative corresponding to $\rho^{(j)}$, then the
symmetric logarithmic derivative of the product state $\rho$ is 
\[
  \cL_p = \sum_j 1 \otimes \cdots \otimes 1 \otimes \cL^{(j)} \otimes \cdots \otimes 1 = \sum_j \cL^{(j)}.
\]
The corresponding Fisher information is
\begin{align*}
  \cI_p &= \Tr \rho_p \cL_p^2 = \sum_j \Tr \rho_p (\cL^{(j)})^2 +
  \sum_{j \ne k} \Tr \rho_p \cL^{(j)}\otimes \cL^{(k)} \nonumber \\ &=  \sum_j \Tr \rho_p (\cL^{(j)})^2 \\ &= \sum_j \cI_p^{(j)} \le n \cI_{\rm bound}.
\end{align*}
The fact that $\Tr \rho_p  \cL^{(j)}\otimes \cL^{(k)} = 0$ for $j \ne
k$ follows almost directly from~\eqref{eq:sld_zero_average}. 

Finally consider an ensemble of states $\rho_e = \sum_j p_j
\rho_j$. To calculate the symmetric logarithmic derivative of an
ensemble it is simpler to start with the corresponding quantum
superposition $\rho_{ex} = \sum_j p_j \rho_j \otimes \ket j \bra j$.
It is easy to check directly that the corresponding symmetric
logarithmic derivative for this superposition is $\cL_{ex} = \sum_j
\cL^{(j)} \otimes \ket j \bra j$.
This gives the quantum Fisher information for the superposition
\[
  \cI_{ex} = \Tr \rho_{ex} \cL_{ex}^2 = \sum_j p_j \cI_{ex}^{(j)} \le \max_j \;\cI_{ex}^{(j)}.
\]
Now, the difference between the quantum ensemble $\rho_e$ and the
quantum superposition $\rho_{ex}$ is simply a trace over the auxiliary
system that marks the particular state of the ensemble. That is, we
forget or lose information. The corresponding Fisher information can
only decrease as a result: $\cI_{e} \le \cI_{ex}$. Formally, this is a
consequence of the monotonicity of the Fisher
information~\cite{petz_monotone_1996}. 

To conclude, we recall that a separable quantum state has the form
$\rho_s = \sum_j p_j \rho_1^{(j)} \otimes \cdots \otimes \rho_N^{(j)}$.
Putting together the bounds for ensembles and product states we
conclude that $\cI_s \le n \cI_{\rm bound}$, or, in other words,
\[
  \delta \hat \phi \ge \frac 1 {\sqrt{n \cI_{\rm bound}}}.
\]

\vspace{.3cm}
\section{Diagrammatic proofs}
\label{sec:diagrams}

Diagrammatically, entangling gates are axiomatically defined by the
following properties, called associativity, isometry,
commutativity, and Frobenius
law~\cite{coeckepaquette:practisingphysicist,coeckeduncan:observables}. 
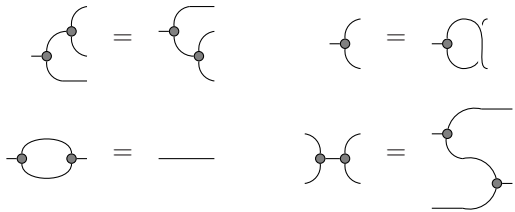
\begin{align*}
  \begin{tikzpicture}[scale=0.33, rotate=-90]
    \node [style=dot] (1) at (0,-1) {};
    \node (5) at (0,1) {};
    \node (0) at (0,-2) {};
    \node (6) at (1,1) {};
    \node (2) at (1,0) {};
    \node (4) at (-2,1) {};
    \node [style=dot] (3) at (-1,0) {};
    \draw (1) to (0);
    \draw [bend right=45]  (1) to (2);
    \draw [bend left=45]  (5) to (3);
    \draw (6) to (2.west);
    \draw [bend right=45]  (4) to (3);
    \draw [bend right=45]  (3) to (1);
  \end{tikzpicture}
  & =
  \begin{tikzpicture}[scale=0.33, rotate=-90]
    \node [style=dot] (1) at (0,-1) {};
    \node (5) at (0,1) {};
    \node (0) at (0,-2) {};
    \node (6) at (-1,1) {};
    \node (2) at (-1,0) {};
    \node (4) at (2,1) {};
    \node [style=dot] (3) at (1,0) {};
    \draw  (1) to (0);
    \draw [bend left=45]  (1) to (2);
    \draw [bend right=45]  (5) to (3);
    \draw  (6) to (2.west);
    \draw [bend left=45]  (4) to (3);
    \draw [bend left=45]  (3) to (1);
  \end{tikzpicture}
  &
  \begin{tikzpicture}[scale=0.33, rotate=-90]
    \node (3) at (-1,2) {};
    \node [style=dot] (1) at (0,1) {};
    \node (0) at (0,0) {};
    \node (2) at (1,2) {};
    \draw [bend right=45]  (1) to (2);
    \draw [bend right=45]  (3) to (1);
    \draw  (1) to (0);
  \end{tikzpicture}
  & =
  \begin{tikzpicture}[scale=0.33,cross/.style={preaction={draw=white,
        -, line width=3pt}}, rotate=-90]
    \node [style=dot] (1) at (0,1) {};
    \node (0) at (0,0) {};
    \node (2) at (1,2) {};
    \node (3) at (-1,2) {};
    \node (4) at (1,3) {};
    \node (5) at (-1,3) {};
    \draw (1) to (0);
    \draw (1) to [bend right] (2) .. controls (4) and (3) .. (5);
    \draw[cross] (1) to [bend left] (3) .. controls (5) and (2) .. (4);
  \end{tikzpicture}
  \\
  \begin{tikzpicture}[scale=0.33, rotate=-90]
    \node (0) at (0,-2) {};
    \node (3) at (0,2) {};
    \node[dot] (2) at (0,1) {};
    \node[dot] (1) at (0,-1) {};
    \draw [bend right=90]  (2) to (1);
    \draw  (2) to (3);
    \draw [bend right=90]  (1) to (2);
    \draw  (0) to (1);
  \end{tikzpicture}
  & =
  \begin{tikzpicture}[scale=0.33, rotate=-90]
    \node (0) at (0,-2) {};
    \node (3) at (0,1) {};
    \draw  (0) to (3);
  \end{tikzpicture}
  &
  \begin{tikzpicture}[scale=0.33, rotate=-90]
    \node (0) at (-1, 2) {};
    \node (1) at (1, 2) {};
    \node [style=dot] (2) at (0, 1) {};
    \node [style=dot] (3) at (0, 0) {};
    \node (4) at (-1, -1) {};
    \node (5) at (1, -1) {};
    \draw (2) to (3);
    \draw[bend left=45] (3) to (5);
    \draw[bend left=45] (4) to (3);
    \draw[bend right=45] (0) to (2);
    \draw[bend right=45] (2) to (1);
  \end{tikzpicture}
  & =
  \begin{tikzpicture}[scale=0.33, rotate=-90]
    \node (0) at (-2, 2) {};
    \node (1) at (1, 2) {};
    \node [style=dot] (2) at (1, 1) {};
    \node (3) at (-2, -0) {};
    \node (4) at (0, -0) {};
    \node (5) at (2, -0) {};
    \node [style=dot] (6) at (-1, -1) {};
    \node (7) at (-1, -2) {};
    \node (8) at (2, -2) {};
    \draw (2) to (1);
    \draw (6) to (7);
    \draw (5) to (8);
    \draw[bend left=45] (2) to (5.west);
    \draw[bend right=45] (6) to (4.west);
    \draw[bend right=45] (3.east) to (6);
    \draw[bend left=45] (4.west) to (2);
    \draw (0) to (3);
  \end{tikzpicture}
\end{align*}
It then follows~\cite{abramskyheunen:hstar} from the structure of
compact categories that there exists a unique morphism 
\mbox{$\vcenter{\hbox{\begin{tikzpicture}[scale=0.33, rotate=90]  
   \node [style=none] (0) at (0, 1) {}; \node [style=dot] (1) at (0, 0) {}; \draw (1) to (0) {};
\end{tikzpicture}}}$} satisfying
\begin{equation}\label{eq:counit}
    \begin{tikzpicture}[scale=0.33, rotate=-90]
     \node [style=none] (0) at (1, 2) {};
     \node [style=dot] (1) at (-1, 1) {};
     \node [style=none] (2) at (1, 1) {};
     \node [style=dot] (3) at (0, 0) {};
     \node [style=none] (4) at (0, -1) {};
     \draw[looseness=1.25, bend left=315] (1) to (3);
     \draw[looseness=1.25, bend right=45] (3) to (2.center);
     \draw (2.center) to (0);
     \draw (4) to (3);   
   \end{tikzpicture} 
  \; = \;
   \begin{tikzpicture}[scale=0.33, rotate=-90]
     \node [style=none] (0) at (0, 2) {};
     \node [style=none] (1) at (0, -1) {};
     \draw (0.center) to (1.center);
   \end{tikzpicture}
\end{equation}

Now Eq.~\eqref{eq:associative} follows immediately from associativity:
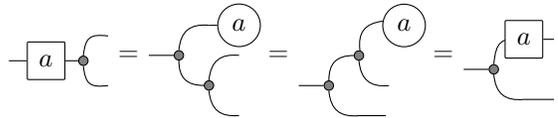
\[\begin{tikzpicture}[scale=0.33, rotate=-90]
        \begin{pgfonlayer}{nodelayer}
        	\node [style=none] (0) at (-1, 1) {};
        	\node [style=none] (1) at (1, 1) {};
        	\node [style=dot] (2) at (0, 0) {};
        	\node [style=box,minimum size = .5cm] (3) at (0, -1.5) {$\;a\;$};
        	\node [style=none] (4) at (0, -3) {};
        \end{pgfonlayer}
        \begin{pgfonlayer}{edgelayer}
        	\draw (2) to (3);
        	\draw[bend right=45, looseness=1.5] (2) to (1.center);
        	\draw (3) to (4.center);
        	\draw[bend right=45, looseness=1.5] (0.center) to (2);
        \end{pgfonlayer}
 \end{tikzpicture}
 =
   \begin{tikzpicture}[scale=0.40, rotate=-90]
     \begin{pgfonlayer}{nodelayer}
     \node [style=none] (0) at (2, 2) {};
     \node [style=none] (1) at (0, 2) {};
     \node [style=circle,draw=black] (2) at (-1, 2) {$a$};
     \node [style=dot] (3) at (1, 1) {};
     \node [style=none] (4) at (-1, 1) {};
     \node [style=dot] (5) at (0, 0) {};
     \node [style=none] (6) at (0, -1) {};
   \end{pgfonlayer}
   \begin{pgfonlayer}{edgelayer}
     \draw (2) to (4.center); \draw[looseness=1.25, bend left=45] (3)
     to (5); \draw[looseness=1.25, bend left=315] (1.center) to (3);
     \draw (5) to (6.center); \draw[looseness=1.25, bend right=315]
     (0.center) to (3); \draw[looseness=1.25, bend right=45]
     (4.center) to (5);
   \end{pgfonlayer}
   \end{tikzpicture} 
   =    
   \begin{tikzpicture}[scale=0.4, rotate=-90]
     \begin{pgfonlayer}{nodelayer}
     \node [style=circle,draw=black] (0) at (-2, 2.5) {$a$};
     \node [style=none] (1) at (0, 2) {};
     \node [style=none] (2) at (1, 2) {};
     \node [style=dot] (3) at (-1, 1) {};
     \node [style=none] (4) at (1, 1) {};
     \node [style=dot] (5) at (0, -0) {};
     \node [style=none] (6) at (0, -1) {};
   \end{pgfonlayer}
   \begin{pgfonlayer}{edgelayer}
     \draw (2) to (4.center);
     \draw[looseness=1.25, bend right=45] (3) to (5);
     \draw[looseness=1.25, bend right=315] (1.center) to (3);
     \draw (5) to (6.center);
     \draw[looseness=1.25, bend left=315] (0) to (3);
     \draw[looseness=1.25, bend left=45] (4.center) to (5);
   \end{pgfonlayer}
   \end{tikzpicture} 
   =
   \begin{tikzpicture}[scale=0.4, rotate=-90]
        \begin{pgfonlayer}{nodelayer}
        	\node [style=none] (0) at (-1, 2) {};
        	\node [style=none] (1) at (1, 2) {};
        	\node [style=box,minimum size = .5cm] (2) at (-1, 1) {$\;a\;$};
        	\node [style=none] (3) at (1, 1) {};
        	\node [style=dot] (4) at (0, 0) {};
        	\node [style=none] (5) at (0, -1) {};
        \end{pgfonlayer}
        \begin{pgfonlayer}{edgelayer}
        	\draw (4) to (5.center);
        	\draw[bend right=45, looseness=1.50] (2.center) to (4);
        	\draw[bend right=45, looseness=1.25] (4) to (3.center);
        	\draw (1.center) to (3.center);
        	\draw (0.center) to (2);
        \end{pgfonlayer}
 \end{tikzpicture}
\]

The transformation between the parallel and sequential
protocols~\eqref{eq:parallelprotocol}
and~\eqref{eq:sequentialprotocol}, as explicitly given in
Eq.~\eqref{eq:useisometry}, follow directly from isometry: 
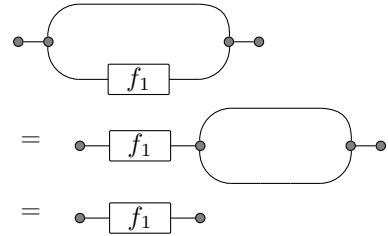
\begin{align*}
      &\begin{tikzpicture}[scale=0.4]
	\begin{pgfonlayer}{nodelayer}
		\node [style=none] (14) at (-2, -1) {};
		\node [style=none] (15) at (2, -1) {};
		\node [style=dot] (21) at (-4, -2.25) {};
		\node [style=dot] (22) at (-3, -2.25) {};
		\node [style=dot] (23) at (3, -2.25) {};
		\node [style=dot] (24) at (4, -2.25) {};
		\node [style=none] (25) at (-2, -3.5) {};
		\node [style=none] (26) at (-1, -3.5) {};
		\node [style=box] (27) at (0, -3.5) {$\;\;f_1\;\;$};
		\node [style=none] (28) at (1, -3.5) {};
		\node [style=none] (29) at (2, -3.5) {};
	\end{pgfonlayer}
	\begin{pgfonlayer}{edgelayer}
                \draw (14.center) to (15.center);
		\draw (25.center) to (26.center);
		\draw[bend left=45] (22) to (14.center);
		\draw[bend right=45] (29.center) to (23);
		\draw (21) to (22);
		\draw[bend left=45, looseness=1.25] (15.center) to (23);
		\draw (26.center) to (28.center);
		\draw[bend left=315] (22) to (25.center);
		\draw (28.center) to (29.center);
		\draw (23) to (24);
	\end{pgfonlayer}
      \end{tikzpicture} \\
     \quad &= \quad
      \begin{tikzpicture}[scale=0.4]
	\begin{pgfonlayer}{nodelayer}
		\node [style=none] (14) at (-2, -1) {};
		\node [style=none] (15) at (1, -1) {};
		\node [style=dot] (21) at (-7, -2.25) {};
		\node [style=dot] (22) at (-3, -2.25) {};
		\node [style=dot] (23) at (2, -2.25) {};
		\node [style=dot] (24) at (3, -2.25) {};
		\node [style=none] (25) at (-2, -3.5) {};
		\node [style=none] (26) at (-1, -3.5) {};
		\node [style=box] (27) at (-5, -2.25) {$\;\;f_1\;\;$};
		\node [style=none] (28) at (0, -3.5) {};
		\node [style=none] (29) at (1, -3.5) {};
	\end{pgfonlayer}
	\begin{pgfonlayer}{edgelayer}
                \draw (14.center) to (15.center);
		\draw (25.center) to (26.center);
		\draw[bend left=45] (22) to (14.center);
		\draw[bend right=45] (29.center) to (23);
		\draw (21) to (22);
		\draw[bend left=45, looseness=1.25] (15.center) to (23);
		\draw (26.center) to (28.center);
		\draw[bend left=315] (22) to (25.center);
		\draw (28.center) to (29.center);
		\draw (23) to (24);
	\end{pgfonlayer}
      \end{tikzpicture} \\ & = \quad
      \begin{tikzpicture}[scale=0.4]
	\begin{pgfonlayer}{nodelayer}
		\node [style=dot] (21) at (-7, -2.25) {};
		\node [style=dot] (22) at (-3, -2.25) {};
		\node [style=box] (27) at (-5, -2.25) {$\;\;f_1\;\;$};
	\end{pgfonlayer}
	\begin{pgfonlayer}{edgelayer}
		\draw (21) to (22);
	\end{pgfonlayer}
      \end{tikzpicture}
\end{align*}

The bijection between the states and operators of
Eqs.~\eqref{eq:abs_operator_state} and~\eqref{eq:abs_phase_point} is
easily seen using the unit property~\eqref{eq:counit}.

Finally, the diagrammatic proof of the commutativity of the linear
operators or quantum maps of Eq.~\eqref{eq:abs_operator_state} 
can now easily be seen to follow from the associative and commutative
properties of the given classical structure. 
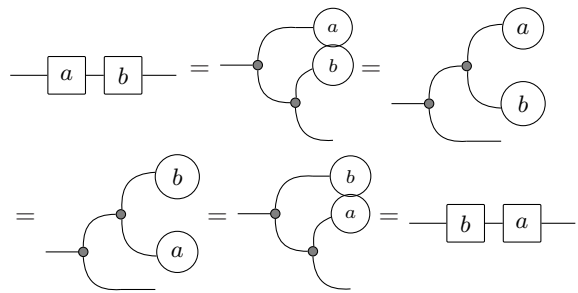
\begin{align*}
  & \begin{tikzpicture}[scale=0.5, font=\small, rotate=-90]
        \begin{pgfonlayer}{nodelayer}
                \node [style=none] (1) at (0,1.5) {};
                \node [style=box,minimum size = .5cm] (2) at (0, 0) {$\;b\;$};
        	\node [style=box,minimum size = .5cm] (3) at (0, -1.5) {$\;a\;$};
        	\node [style=none] (4) at (0, -3) {};
        \end{pgfonlayer}
        \begin{pgfonlayer}{edgelayer}
        	\draw (2) to (1);
        	\draw (3) to (4.center);
                \draw (3) to (2);
        \end{pgfonlayer}
 \end{tikzpicture}
  =
   \begin{tikzpicture}[scale=0.5, font=\scriptsize, rotate=-90]
     \begin{pgfonlayer}{nodelayer}
     \node [style=none] (0) at (2, 2) {};
     \node [style=circle,draw=black] (1) at (0, 2) {$b$};
     \node [style=circle,draw=black] (2) at (-1, 2) {$a$};
     \node [style=dot] (3) at (1, 1) {};
     \node [style=none] (4) at (-1, 1) {};
     \node [style=dot] (5) at (0, 0) {};
     \node [style=none] (6) at (0, -1) {};
   \end{pgfonlayer}
   \begin{pgfonlayer}{edgelayer}
     \draw (2) to (4.center); 
     \draw[looseness=1.25, bend left=45] (3) to (5); 
     \draw[looseness=1.25, bend left=325] (1) to (3);
     \draw (5) to (6.center); 
     \draw[looseness=1.25, bend right=315] (0.center) to (3); 
     \draw[looseness=1.25, bend right=45] (4.center) to (5);
   \end{pgfonlayer}
   \end{tikzpicture} 
   =
   \begin{tikzpicture}[scale=0.5, font=\small, rotate=-90]
     \begin{pgfonlayer}{nodelayer}
     \node [style=circle,draw=black] (0) at (-2, 2.5) {$a$};
     \node [style=circle,draw=black] (1) at (0, 2.5) {$b$};
     \node [style=none] (2) at (1, 2) {};
     \node [style=dot] (3) at (-1, 1) {};
     \node [style=none] (4) at (1, 1) {};
     \node [style=dot] (5) at (0, -0) {};
     \node [style=none] (6) at (0, -1) {};
   \end{pgfonlayer}
   \begin{pgfonlayer}{edgelayer}
     \draw (2) to (4.center);
     \draw[looseness=1.25, bend right=45] (3) to (5);
     \draw[looseness=1.25, bend right=315] (1) to (3);
     \draw (5) to (6.center);
     \draw[looseness=1.25, bend left=315] (0) to (3);
     \draw[looseness=1.25, bend left=45] (4.center) to (5);
   \end{pgfonlayer}
   \end{tikzpicture} 
   \\ & =
  \begin{tikzpicture}[scale=0.5, font=\small, rotate=-90]
     \begin{pgfonlayer}{nodelayer}
     \node [style=circle,draw=black] (0) at (-2, 2.5) {$b$};
     \node [style=circle,draw=black] (1) at (0, 2.5) {$a$};
     \node [style=none] (2) at (1, 2) {};
     \node [style=dot] (3) at (-1, 1) {};
     \node [style=none] (4) at (1, 1) {};
     \node [style=dot] (5) at (0, -0) {};
     \node [style=none] (6) at (0, -1) {};
   \end{pgfonlayer}
   \begin{pgfonlayer}{edgelayer}
     \draw (2) to (4.center);
     \draw[looseness=1.25, bend right=45] (3) to (5);
     \draw[looseness=1.25, bend right=315] (1) to (3);
     \draw (5) to (6.center);
     \draw[looseness=1.25, bend left=315] (0) to (3);
     \draw[looseness=1.25, bend left=45] (4.center) to (5);
   \end{pgfonlayer}
   \end{tikzpicture} 
  =
   \begin{tikzpicture}[scale=0.5, font=\scriptsize, rotate=-90]
     \begin{pgfonlayer}{nodelayer}
     \node [style=none] (0) at (2, 2) {};
     \node [style=circle,draw=black] (1) at (0, 2) {$a$};
     \node [style=circle,draw=black] (2) at (-1, 2) {$b$};
     \node [style=dot] (3) at (1, 1) {};
     \node [style=none] (4) at (-1, 1) {};
     \node [style=dot] (5) at (0, 0) {};
     \node [style=none] (6) at (0, -1) {};
   \end{pgfonlayer}
   \begin{pgfonlayer}{edgelayer}
     \draw (2) to (4.center); 
     \draw[looseness=1.25, bend left=45] (3) to (5); 
     \draw[looseness=1.25, bend left=325] (1) to (3);
     \draw (5) to (6.center); 
     \draw[looseness=1.25, bend right=315] (0.center) to (3); 
     \draw[looseness=1.25, bend right=45] (4.center) to (5);
   \end{pgfonlayer}
   \end{tikzpicture} 
   =
  \begin{tikzpicture}[scale=0.5, font=\small, rotate=-90]
        \begin{pgfonlayer}{nodelayer}
                \node [style=none] (1) at (0,1.5) {};
                \node [style=box,minimum size = .5cm] (2) at (0, 0) {$\;a\;$};
        	\node [style=box,minimum size = .5cm] (3) at (0, -1.5) {$\;b\;$};
        	\node [style=none] (4) at (0, -3) {};
        \end{pgfonlayer}
        \begin{pgfonlayer}{edgelayer}
        	\draw (2) to (1);
        	\draw (3) to (4.center);
                \draw (3) to (2);
        \end{pgfonlayer}
 \end{tikzpicture}
\end{align*}
These proofs only rely on the four axioms of classical structures, and
hence hold in any dagger compact category rather than just for
finite-dimensional Hilbert spaces with either linear or quantum maps.

\bibliography{arxiv-sequentializable}

\end{document}